\documentclass[pra,reprint]{revtex4-1}

\usepackage[ngerman, english]{babel}
\usepackage{graphicx}
\usepackage{dcolumn}%
\usepackage{bm}%
\usepackage{geometry}
\usepackage{array}
\usepackage{amsthm}
\usepackage{amstext}
\usepackage{amsmath,amssymb}
\usepackage{nicefrac,subfigure}
\usepackage{amsopn}

\usepackage{color}

\def\({\left(} \def\){\right)}  \def\lw{\left\langle} \def\rw{\right\rangle}

\let\ph=\varphi 

\def\e{{\rm e}}
\def\C{\mathbb C}
\def\N{\mathbb N}
\def\R{\mathbb R}
\def\Rd{{\mathbb{R}^d}}
\def\Rdd{{\mathbb{R}^{2d}}}
\def\wh{\widehat}

\def\H{{\mathcal H}}
\def\W{{\mathcal W}}

\begin{document}

\draft

\title{Quantum expectations via spectrograms}
\author{Johannes Keller}
\email[]{keller@ma.tum.de}
\author{Caroline Lasser}
\email[]{classer@ma.tum.de}
\affiliation{Zentrum Mathematik, Technische Universit\"at M\"unchen, 80290 M\"unchen, Germany}

\author{Tomoki Ohsawa}
\email[]{tomoki@utdallas.edu}
\affiliation{Department of Mathematical Sciences, The University of Texas at Dallas, Richardson, TX 75080-3021}

\date{\today}

\begin{abstract}
We discuss a new phase space method for the computation of quantum expectation values in the high frequency regime. 
Instead of representing a wavefunction by its Wigner function, which typically attains negative values, we define a new phase space density by adding a first-order Hermite spectrogram term as a correction to the Husimi function.
The new phase space density yields accurate approximations of the quantum expectation values as well as allows numerical sampling from non-negative densities. We illustrate the new method by numerical experiments in up to $128$ dimensions.
\end{abstract}


\maketitle


\section{Introduction}

Wigner functions are phase space functions that represent quantum states~\cite{HWSW84}. Among their important properties is the exact 
correspondence for expectation values via phase space integration. However, Wigner functions in general attain negative values. 
This phenomenon of ``negative probabilities'' also poses numerical problems when discretizing the phase space integrals by Monte-Carlo methods.

Spectrograms are a large class of non-negative phase space functions that result from coarse-graining the Wigner function 
by convolution with another Wigner function~\cite{Flandrin15}. The most prominent spectrogram is the Husimi function~\cite{H40}, which is obtained by convolving 
with the Wigner function of a Gaussian. The drawback of the coarse-graining is the loss of the exact correspondence for expectation values.

Following the ideas of~\cite{KLO15}, we add first order Hermite spectrograms to the Husimi function.  The resulting new phase 
space density is a linear combination of probability densities that approximates the Wigner function more accurately than the Husimi function, particularly in calculating the expectation values of observables.

We proceed as follows. After a brief review of the relationship between expectation values and Wigner and Husimi functions as well as of spectrograms in \S\ref{sec:exp}, 
we construct the new density in \S\ref{sec:dens} and present supporting numerical experiments in \S\ref{sec:num}.
\section{Quantum Expectations}\label{sec:exp}

\subsection{High frequency wavefunctions}

Our goal is to compute quantum expectation values
of observables $\wh A$,
\begin{equation}\label{eq:quantum-exp}
\lw\psi | \wh A |\psi\rw =:\lw \wh A\rw_\psi
\end{equation}
for wavefunctions $\psi:\Rd \to \C$ that are highly oscillatory with frequencies of the size $O(h^{-1})$, where 
\[
0<h\ll 1
\]
is a small parameter. In the context of the Born--Oppenheimer approximation, $h$ is square root of the ratio of electronic 
versus average nuclear mass and typically ranges between $10^{-3}$ and $10^{-1}$.

We assume that the observable $\wh A$ arises as the Weyl quantization of its
classical phase space counterpart $A:\Rdd \to \R$; see e.g.~\cite{C08}. Then, expectation values
can be computed via the integral formula
\begin{equation}\label{eq:weyl_expectation}
\lw \wh A\rw_\psi = \int_\Rdd A(z) \W_\psi(z)dz,
\end{equation}
where the Wigner function $\W_\psi:\Rdd \to \R$ is the phase space symbol of the quantum state,
\[
\widehat\W_\psi  \propto |\psi\rangle\langle \psi|,
\]
see e.g.~\cite{C08,C89,HWSW84,Ber77} or the monograph~\cite[\S 3]{Schl11}.

\subsection{Phase space densities}

\begin{figure}[h!]
\includegraphics[width = 0.45\textwidth]{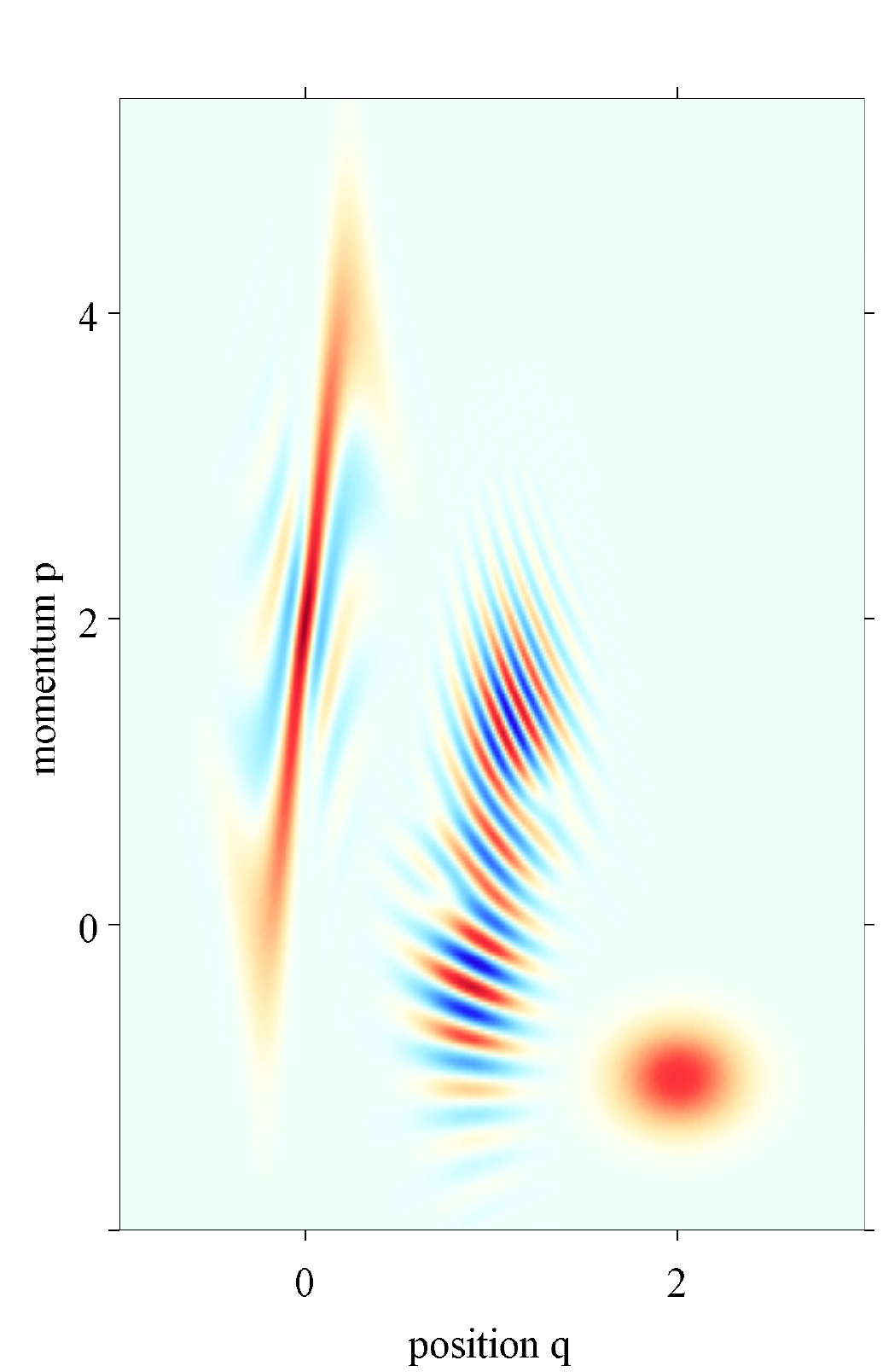}
\caption{ \label{fig:wigner} Contour plot of the Wigner function of a one-dimensional superposition of a Gaussian state (lower right)
and a WKB type delocalized state (left). Blue coloring is used for negative values.}
\end{figure}

Apart from Gaussian states, Wigner functions attain negative values and usually exhibit strong oscillations; see the illustration in Figure~\ref{fig:wigner}. 
Hence it is difficult to sample phase space points with respect to $\W_\psi$ for the Monte Carlo quadrature of the integral~\eqref{eq:weyl_expectation}.
Our goal is to replace $\W_\psi$ by a new phase space density with the following properties:
\begin{itemize}
\item the new density is built from smooth probability densities, so that~\eqref{eq:weyl_expectation} is amenable to positive 
sampling strategies;
\item the new density approximates the Wigner function, so that \eqref{eq:weyl_expectation} is exact for quadratic observables and holds approximately with a small error in general.
\end{itemize}
 
It is well-known that Husimi functions and spectrograms
are analytic probability densities, but only provide exact expectation values for linear observables.
However, an appropriate linear combination of Hermite spectrograms results in a
phase space density that fulfills both our requirements.

\subsection{Spectrograms and Husimi functions}\label{sec:Spec_and_Hus}

Non-negative phase space representations can be obtained by coarse-graining of the Wigner function.
The spectrogram $S^\phi_\psi:\Rdd \to [0,\infty)$ of a wavefunction $\psi$ with respect to a smooth window function
$\phi:\Rd \to \C$ is defined as the convolution
\begin{align*}
S^\phi_\psi(z) &:= (\W_\psi * \W_\phi)(z)\\
&= \int_\Rdd \W_\psi(w)\W_\phi(z-w) dw.
\end{align*}
Whenever $\psi$ and $\phi$ are $L^2$-normalized,
$S^\phi_\psi$ is a smooth probability density on the phase space.
Spectrograms are widely used in time-frequency analysis; see e.g.~the introduction of~\cite{Flandrin15}.

The most popular spectrogram is the Husimi function, 
\begin{equation}
  \label{eq:Husimi}
  \H_\psi := S^{g_{0}}_\psi = \W_{\psi} * \W_{g_{0}},
\end{equation}
see e.g.~\cite{H40,T86}, where the window is a Gaussian function 
\begin{equation*}
  g_{0}(x) := (\pi h)^{-d/4}\exp(-\tfrac{1}{2h}|x|^2).
\end{equation*}

Replacing the Wigner function by the Husimi function
in equation~\eqref{eq:weyl_expectation}, we obtain 
\begin{equation}\label{eq:husimi_method}
\lw \wh A\rw_\psi = \int_\Rdd A(z) \H_\psi(z)dz + O(h),
\end{equation}
where the error term vanishes for linear observables $\wh A$; see also Figure~\ref{fig:accuracy}.

\section{A new phase space density}\label{sec:dens}

\subsection{Combining Hermite spectrograms}\label{sec:density}

\begin{figure}[h!]
\includegraphics[width = 0.45\textwidth]{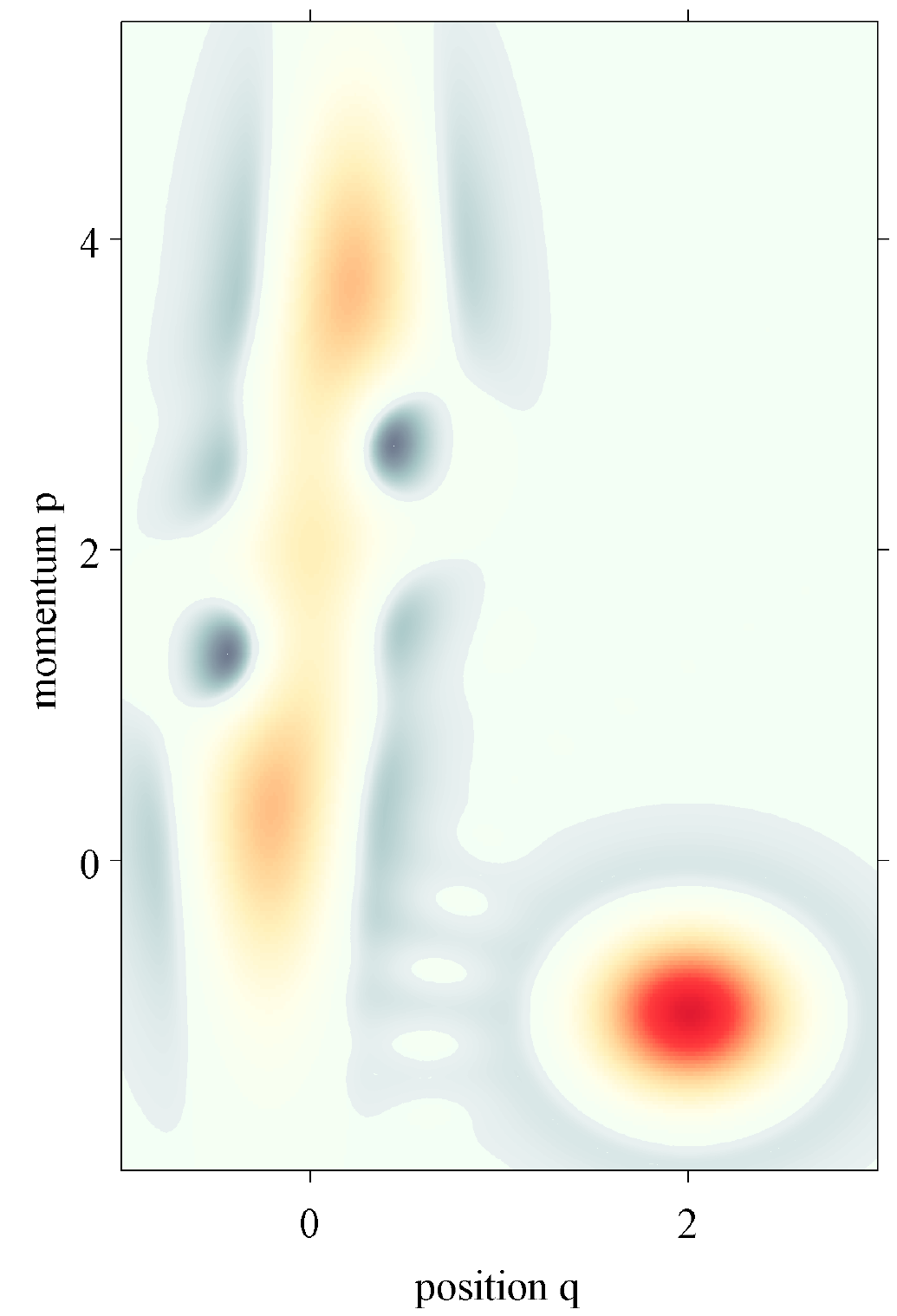}
\caption{ \label{fig:mu} Contour plot of the density $\mu_\psi$ for the same state as in Figure~\ref{fig:wigner}.
Grey coloring is used to highlight regions of prominent negative values.}
\end{figure}

We Taylor expand the Gaussian in the convolution defining the Husimi function
\[
\H_\psi(z) = (\pi h)^{-d} \int_\Rdd  \W_\psi(w) \e^{-|z-w|^2/h} dw
\]
around $z$, and obtain the asymptotic expansion
\begin{equation*}
  \W_\psi(z) = \H_\psi(z) - \frac{h}{4} \Delta\H_\psi(z) + O(h^2).
\end{equation*}
Substituting this expression into \eqref{eq:weyl_expectation} yields
\begin{align*}
\lw \wh A\rw_\psi 
&=  \int_\Rdd  \!\!\!\!A(z)(1-\tfrac{h}4 \Delta )\H_\psi(z)dz + O(h^2),
\end{align*}
where the error depends on fourth and higher order derivatives of $A$.
We take a closer look at the Laplacian $\Delta\H_\psi$. Consider the rescaled
first order Hermite functions
\[
\ph_{j}(x) = (\pi h)^{-d/4} \sqrt{\tfrac2h}x_j \exp\(-\tfrac{1}{2h}|x|^2\),
\]
for $j=1,\hdots, d$. A direct calculation provides a first order Laguerre 
function in $|z|^2/h$, 
\[
\W_{\ph_{j}}(z) = -(\pi h)^{-d}(1-\tfrac2h|z_j|^2)\e^{-|z|^2/h}
\]
for $z=(z_1,\ldots,z_d)\in \Rdd$; see also \cite{Gro46}. Thus,
\begin{align*}
\Delta \W_{g_0}(z) &=\tfrac{2}h\sum_{j=1}^d \W_{\ph_{j}}(z) - \tfrac{2d}h\W_{g_0}(z),
\end{align*}
and then we conclude from \eqref{eq:Husimi} that
\[
\Delta \H_{\psi} = \W_{\psi}*\Delta\W_{g_0} = \tfrac2h \sum_{j=1}^d S_\psi^{\ph_j} - \tfrac{2d}h \H_\psi.
\]
Now we define the new phase space density $\mu_\psi:\Rdd\to\R$ by
\begin{equation}
\mu_\psi(z) := (1+\tfrac{d}2)\H_\psi(z) - \tfrac{1}2\sum_{j=1}^d S_\psi^{\ph_j}(z),
\end{equation}
as a linear combination of the two smooth probability densities $\H_\psi$ and $\frac{1}d\sum_{j=1}^d S_\psi^{\ph_j}$;
see Figure~\ref{fig:mu} for an example. The new density satisfies
\[
\int_{\Rdd} \mu_\psi(z) dz = \langle\psi|\psi\rangle^2
\]
and provides quantum expectations as
\begin{equation}\label{eq:mu_second_order}
\lw \wh A\rw_\psi  =\int_\Rdd  A(z)\mu_\psi(z)dz + O(h^2).
\end{equation}
The error term in~\eqref{eq:mu_second_order} vanishes whenever $A$ is a polynomial of degree less or equal to three. 
By including higher order Hermite spectrograms, 
one can derive similar densities that give exact expectation values for higher order polynomial observables;
see~\cite{K16}.

\subsection{Explicit formulas for $\mu_\psi$}\label{sec:examples}

One can find formulas for $\mu_\psi$ when the state is a Gaussian wavepacket, $\psi = g_z$ ,
\begin{align*}
g_{z}(x) :=& (\pi h)^{-d/4} \exp(\tfrac ih p\cdot(x-\tfrac12q))\\
& \times \exp(-\tfrac{1}{2h}|x-q|^2)
\end{align*}
centered in $z=(q,p)\in\Rdd$, 
a Gaussian superposition, $\psi = g_{z_1} + g_{z_2}$ ,
or a harmonic oscillator eigenstate, $\psi = \ph_k$, $k\in\N^d$; see \cite{KLO15}.

\begin{figure}[h!]
\includegraphics[width = 0.48\textwidth]{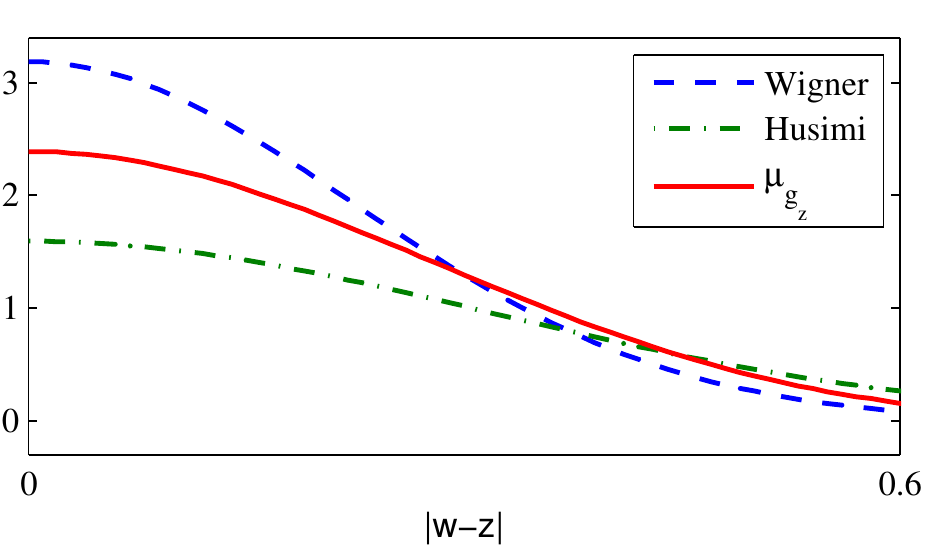}
\caption{ \label{fig:decay} Section of the Wigner function $\W_{g_z}(w)$, the Husimi function $\H_{g_z}(w)$, and
 the density $\mu_{g_z}(w)$ in dependence of $|w-z|$, where $h = 10^{-1}$.}
\end{figure}

For instance, a Gaussian wave packet gives rise to the density
\[
\mu_{g_z}(w) =  (2\pi  h)^{-d} \!\left(1+\tfrac{d}{2}-\tfrac{|w-z|^2}{4h}\right) \! \e^{-|w-z|^2/2h} .
\]
Figure~\ref{fig:decay} illustrates that --- due to the polynomial prefactor ---
the density~$\mu_{g_z}$ lies between the Wigner and the Husimi
function.

For Gaussian superpositions one has
\begin{equation}\label{eq:super}
\mu_{g_{z_1} + g_{z_2}} = \mu_{g_{z_1}}  + \mu_{g_{z_2}} + \e^{-|z_1 - z_2|^2/8h} c_{1,2}
\end{equation}
where $c_{1,2}$ is an oscillatory interference term; see Appendix \ref{app}.
Due to the exponentially small prefactor $\e^{-|z_1 - z_2|^2/8h}$ the cross term may be 
neglected for numerical computations whenever the two centers $z_1$ and $z_2$ are sufficiently apart from each other.

\section{Numerical experiments}\label{sec:num}

\subsection{Accuracy: Gaussian superposition}

\begin{figure}[h!]
\includegraphics[width = 0.48\textwidth]{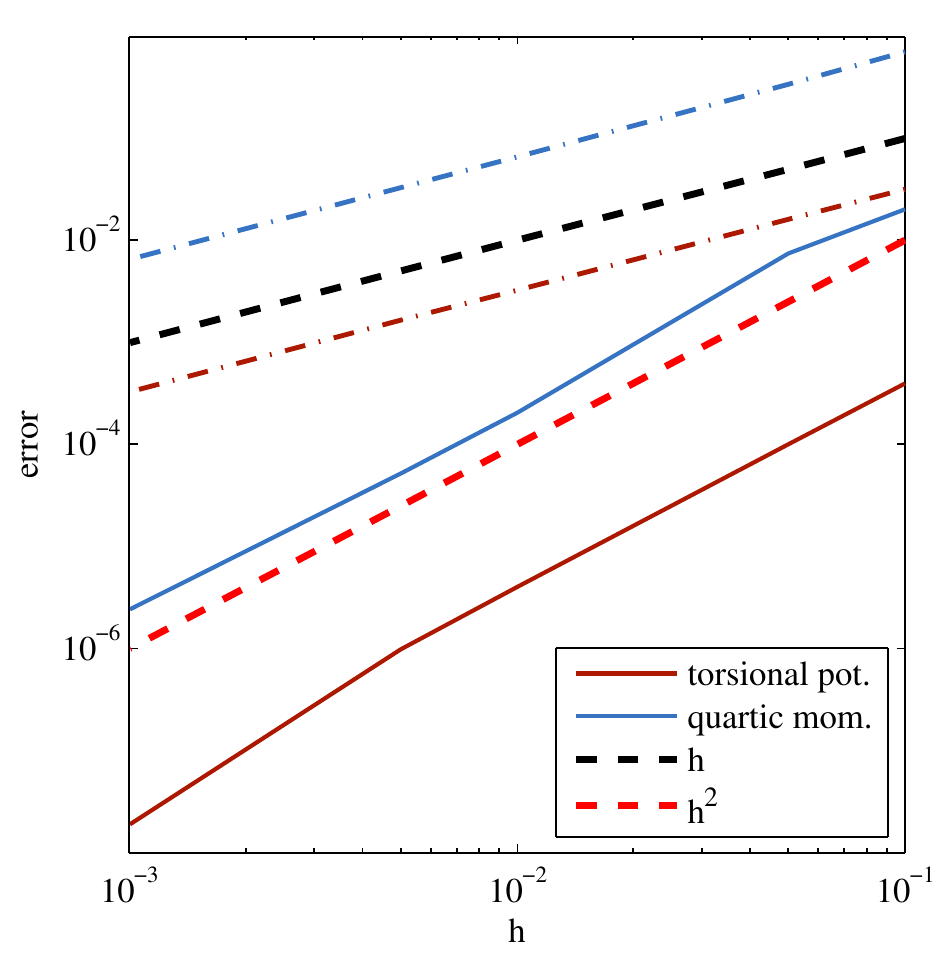}
\caption{ \label{fig:accuracy} Approximation errors of expectation values
for the spectrogram method~\eqref{eq:mu_second_order} (solid lines) and  the Husimi method~\eqref{eq:husimi_method} (dashed dotted lines).
}
\end{figure}

In a first set of experiments we implement a Quasi-Monte Carlo discretization 
 of our new method~\eqref{eq:mu_second_order} in two dimensions with Halton points; see also~\cite{KLO15}.
We test the accuracy for a Gaussian superposition $\psi=g_{z_1} + g_{z_2}$, 
$z_1 = (-1,1, 1, 1),  z_2 = (0, 1, -1, -\frac12)$, with varying
values of~$h$ and the following observables:
\begin{itemize}
\item  torsional  potential: $ 2-\cos(\wh q_1)-\cos(\wh q_2)$
\item  quartic momentum: $ |\wh p|^4$
\end{itemize}
We  compare the outcome with reference values from grid based quadrature and do the same for the 
Husimi method~\eqref{eq:husimi_method}; see Figure~\ref{fig:accuracy}.
Since both observables have non-vanishing derivatives of degree four,
the errors of the spectrogram method are of size $O(h^2)$.
In contrast, for the Husimi method~\eqref{eq:husimi_method},
the expectation values are  approximated with $O(h)$ accuracy only.

\subsection{High dimensions: Henon--Heiles}

In a second set of experiments we show the applicability of the new method in moderately
high dimensions. We consider the Henon--Heiles Hamiltonian
\begin{align}
\widehat H &=\nonumber -\tfrac{h^2}2 \Delta  + \sum_{j=1}^d \tfrac{q_j^2}2   +  \alpha \sum_{j=1}^{d-1}(q_j^2q_{j+1} - \tfrac{q_{j+1}^3}3 )\\
& \label{eq:henon-ham}=: \tfrac12|\wh p|^2  + V_d(q)
\end{align}
with $\alpha = 1.8436$ and $h=0.0037$ in dimensions $d=2,\hdots,128$, and
 a Gaussian state $\psi=g_{z}$ centered in $z=(q,p)$ with $q_1=\hdots =q_d =  0.3645$ and nonzero momentum
$ p_1 = 1, p_2=\hdots =p_d=0$.

\begin{figure}[h!]
\includegraphics[width = 0.48\textwidth]{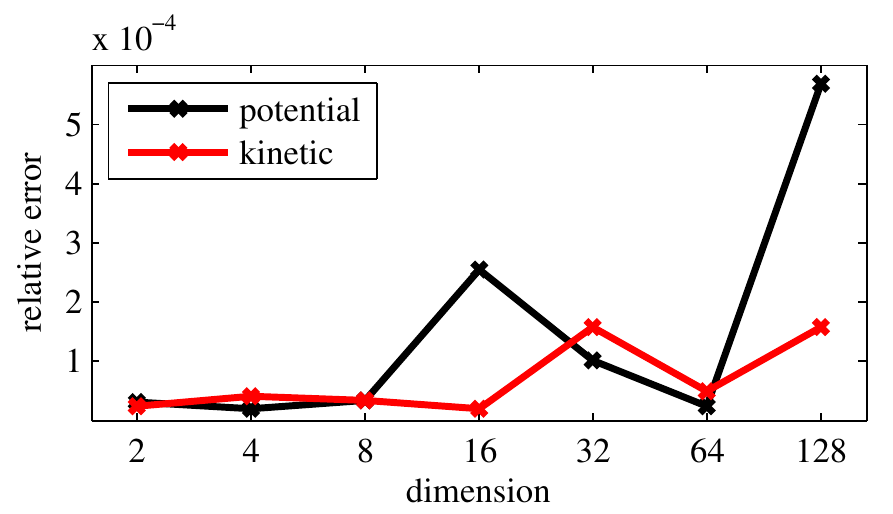}
\caption{ \label{fig:halton} Relative Henon--Heiles energy errors in dimensions $d=2,\hdots ,128$ with $10^8$ Monte-Carlo points.
}
\end{figure}

The Hamiltonian $\widehat H $ and the state~$\psi$ constitute a benchmark system for time-dependent propagation methods;
see e.g.~\cite{MMC90,WMM01}. Since the potential function $V_d$ is a polynomial of degree three, the spectrogram method~\eqref{eq:mu_second_order} 
gives the exact kinetic and potential energies, and all errors are sampling errors, see~Figure~\ref{fig:halton}.

\section{Conclusion}

We have suggested a new phase space density for computing quantum expectation values by taking a linear combination of spectrograms.
The new phase space density not only remedies the difficulty in sampling with the Wigner function but also provides a higher-order approximation of observables than the Husimi function in the high frequency regime.

\appendix
\section{Cross terms}\label{app}
For the Gaussian superposition $\psi=g_{z_1}+g_{z_2}$, both the Wigner function and the new density contain an oscillatory cross term that localizes 
around the arithmetic mean $z_+ = \tfrac12(z_1+z_2)$.
For the Wigner function, we have
\[
\W_\psi = \W_{g_{z_1}} + \W_{g_{z_2}} + 2\gamma_{1,2}
\]
with
\begin{align*}
\gamma_{1,2}(z) &= (\pi h)^{-d} \exp(-\tfrac1h |z-z_+|^2)\\
& \quad\times\cos(\tfrac1h (z_1-z_2)\cdot\Omega z),
\end{align*}
where
\[
\Omega := \begin{pmatrix}0 & {\rm Id}_{d\times d}\\ -{\rm Id}_{d\times d} & 0\end{pmatrix}\in\R^{2d\times 2d}.
\]

For the new density, we obtain equation~\eqref{eq:super} with cross term
\begin{align*}
&c_{1,2}(z) = (2\pi h)^{-d} \exp(-\tfrac{1}{2h} |z-z_+|^2)\\
& \times\left((z-z_1)\cdot(z-z_2)\cos(\tfrac{1}{2h} (z_1-z_2)\cdot\Omega z)\right.\\
& \quad \left.-(z-z_1)\cdot\Omega (z-z_2)\sin(\tfrac{1}{2h} (z_1-z_2)\cdot\Omega z)\right)
\end{align*}
In comparison to the Wigner function, the cross term of the new density is damped by the small constant $\e^{-|z_1-z_2|^2/8h}$.


\begin{thebibliography}{10}%
\makeatletter
\providecommand \@ifxundefined [1]{%
 \ifx #1\undefined \expandafter \@firstoftwo
 \else \expandafter \@secondoftwo
\fi
}%
\providecommand \@ifnum [1]{%
 \ifnum #1\expandafter \@firstoftwo
 \else \expandafter \@secondoftwo
\fi
}%
\providecommand \enquote [1]{``#1''}%
\providecommand \bibnamefont  [1]{#1}%
\providecommand \bibfnamefont [1]{#1}%
\providecommand \citenamefont [1]{#1}%
\providecommand\href[0]{\@sanitize\@href}%
\providecommand\@href[1]{\endgroup\@@startlink{#1}\endgroup\@@href}%
\providecommand\@@href[1]{#1\@@endlink}%
\providecommand \@sanitize [0]{\begingroup\catcode`\&12\catcode`\#12\relax}%
\@ifxundefined \pdfoutput {\@firstoftwo}{%
 \@ifnum{\z@=\pdfoutput}{\@firstoftwo}{\@secondoftwo}%
}{%
 \providecommand\@@startlink[1]{\leavevmode}%
 \providecommand\@@endlink[0]{}%
}{%
 \providecommand\@@startlink[1]{%
  \leavevmode
  \pdfstartlink
   attr{/Border[0 0 1 ]/H/I/C[0 1 1]}%
   user{/Subtype/Link/A<</Type/Action/S/URI/URI(#1)>>}%
  \relax
 }%
 \providecommand\@@endlink[0]{\pdfendlink}%
}%
\providecommand \url  [0]{\begingroup\@sanitize \@url }%
\providecommand \@url [1]{\endgroup\@href {#1}{\urlprefix}}%
\providecommand \urlprefix [0]{URL }%
\providecommand \Eprint[0]{\href }%
\@ifxundefined \urlstyle {%
  \providecommand \doi [1]{doi:\discretionary{}{}{}#1}%
}{%
  \providecommand \doi [0]{doi:\discretionary{}{}{}\begingroup
  \urlstyle{rm}\Url }%
}%
\providecommand \doibase [0]{http://dx.doi.org/}%
\providecommand \Doi[1]{\href{\doibase#1}}%
\providecommand \bibAnnote [3]{%
  \BibitemShut{#1}%
  \begin{quotation}\noindent
    \textsc{Key:}\ #2\\\textsc{Annotation:}\ #3%
  \end{quotation}%
}%
\providecommand \bibAnnoteFile [2]{%
  \IfFileExists{#2}{\bibAnnote {#1} {#2} {\input{#2}}}{}%
}%
\providecommand \typeout [0]{\immediate \write \m@ne }%
\providecommand \selectlanguage [0]{\@gobble}%
\providecommand \bibinfo [0]{\@secondoftwo}%
\providecommand \bibfield [0]{\@secondoftwo}%
\providecommand \translation [1]{[#1]}%
\providecommand \BibitemOpen[0]{}%
\providecommand \bibitemStop [0]{}%
\providecommand \bibitemNoStop [0]{.\EOS\space}%
\providecommand \EOS [0]{\spacefactor3000\relax}%
\providecommand \BibitemShut [1]{\csname bibitem#1\endcsname}%
\bibitem{HWSW84}%
  \BibitemOpen
  \bibfield{author}{%
  \bibinfo {author} {\bibfnamefont{M.}~\bibnamefont{Hillery}}, \bibinfo
  {author} {\bibfnamefont{R.}~\bibnamefont{O'Connell}}, \bibinfo {author}
  {\bibfnamefont{M.}~\bibnamefont{Scully}},\ and\ \bibinfo {author}
  {\bibfnamefont{E.}~\bibnamefont{Wigner}},\ }%
  \bibfield{journal}{%
  \Doi{http://dx.doi.org/10.1016/0370-1573(84)90160-1}{\bibinfo {journal}
  {Physics Reports}}\ }%
  \textbf{\bibinfo {volume} {106}},\ \bibinfo {pages} {121 } (\bibinfo {year}
  {1984}),\ ISSN \bibinfo {issn} {0370-1573}%
  \bibAnnoteFile{NoStop}{HWSW84}%
\bibitem{Flandrin15}%
  \BibitemOpen
  \bibfield{author}{%
  \bibinfo {author} {\bibfnamefont{P.}~\bibnamefont{Flandrin}},\ }%
  \bibfield{journal}{%
  \Doi{10.1109/LSP.2015.2463093}{\bibinfo {journal} {IEEE Signal Processing
  Letters}}\ }%
  \textbf{\bibinfo {volume} {22}},\ \bibinfo {pages} {2137} (\bibinfo {month}
  {Nov}\ \bibinfo {year} {2015}),\ ISSN \bibinfo {issn} {1070-9908}%
  \bibAnnoteFile{NoStop}{Flandrin15}%
\bibitem{H40}%
  \BibitemOpen
  \bibfield{author}{%
  \bibinfo {author} {\bibfnamefont{K.}~\bibnamefont{Husimi}},\ }%
  \bibfield{journal}{%
  \bibinfo {journal} {Proceedings of the Physico-Mathematical Society of Japan.
  3rd Series}\ }%
  \textbf{\bibinfo {volume} {22}},\ \bibinfo {pages} {264} (\bibinfo {year}
  {1940})%
  \bibAnnoteFile{NoStop}{H40}%
\bibitem{KLO15}%
  \BibitemOpen
  \bibfield{author}{%
  \bibinfo {author} {\bibfnamefont{J.}~\bibnamefont{Keller}}, \bibinfo {author}
  {\bibfnamefont{C.}~\bibnamefont{Lasser}},\ and\ \bibinfo {author}
  {\bibfnamefont{T.}~\bibnamefont{Ohsawa}},\ }%
  \bibfield{journal}{%
  \Doi{10.1137/15M1028388}{\bibinfo {journal} {SIAM J. Math. Anal.}}\ }%
  \textbf{\bibinfo {volume} {48}},\ \bibinfo {pages} {513} (\bibinfo {year}
  {2016})%
  \bibAnnoteFile{NoStop}{KLO15}%
\bibitem{C08}%
  \BibitemOpen
  \bibfield{author}{%
  \bibinfo {author} {\bibfnamefont{W.~B.}\ \bibnamefont{Case}},\ }%
  \bibfield{journal}{%
  \bibinfo {journal} {American Journal of Physics}\ }%
  \textbf{\bibinfo {volume} {76}},\ \bibinfo {pages} {937} (\bibinfo {year}
  {2008})%
  \bibAnnoteFile{NoStop}{C08}%
\bibitem{C89}%
  \BibitemOpen
  \bibfield{author}{%
  \bibinfo {author} {\bibfnamefont{L.}~\bibnamefont{Cohen}},\ }%
  \bibfield{journal}{%
  \bibinfo {journal} {Proceedings of the IEEE}\ }%
  \textbf{\bibinfo {volume} {77}},\ \bibinfo {pages} {941} (\bibinfo {year}
  {1989})%
  \bibAnnoteFile{NoStop}{C89}%
\bibitem{Ber77}%
  \BibitemOpen
  \bibfield{author}{%
  \bibinfo {author} {\bibfnamefont{M.~V.}\ \bibnamefont{Berry}},\ }%
  \bibfield{journal}{%
  \bibinfo {journal} {Philosophical Transactions of the Royal Society of London
  A: Mathematical, Physical and Engineering Sciences}\ }%
  \textbf{\bibinfo {volume} {287}},\ \bibinfo {pages} {237} (\bibinfo {year}
  {1977})%
  \bibAnnoteFile{NoStop}{Ber77}%
\bibitem{Schl11}%
  \BibitemOpen
  \bibfield{author}{%
  \bibinfo {author} {\bibfnamefont{W.~P.}\ \bibnamefont{Schleich}},\ }%
  \emph{\bibinfo {title} {Quantum optics in phase space}}\ (\bibinfo
  {publisher} {John Wiley \& Sons},\ \bibinfo {year} {2011})%
  \bibAnnoteFile{NoStop}{Schl11}%
\bibitem{T86}%
  \BibitemOpen
  \bibfield{author}{%
  \bibinfo {author} {\bibfnamefont{K.}~\bibnamefont{Takahashi}},\ }%
  \bibfield{journal}{%
  \Doi{10.1143/JPSJ.55.762}{\bibinfo {journal} {Journal of the Physical Society
  of Japan}}\ }%
  \textbf{\bibinfo {volume} {55}},\ \bibinfo {pages} {762} (\bibinfo {year}
  {1986})%
  \bibAnnoteFile{NoStop}{T86}%
\bibitem{Gro46}%
  \BibitemOpen
  \bibfield{author}{%
  \bibinfo {author} {\bibfnamefont{H.}~\bibnamefont{Groenewold}},\ }%
  \bibfield{journal}{%
  \bibinfo {journal} {Physica}\ }%
  \textbf{\bibinfo {volume} {12}},\ \bibinfo {pages} {405} (\bibinfo {year}
  {1946})%
  \bibAnnoteFile{NoStop}{Gro46}%
\bibitem{K16}%
  \BibitemOpen
  \bibfield{author}{%
  \bibinfo {author} {\bibfnamefont{J.}~\bibnamefont{Keller}},\ }%
  \enquote{\bibinfo {title} {The spectrogram expansion of {W}igner
  functions},}\  (\bibinfo {year} {2016}),\ \bibinfo {note} {in preparation}%
  \bibAnnoteFile{NoStop}{K16}%
\bibitem{MMC90}%
  \BibitemOpen
  \bibfield{author}{%
  \bibinfo {author} {\bibfnamefont{H.-D.}\ \bibnamefont{Meyer}}, \bibinfo
  {author} {\bibfnamefont{U.}~\bibnamefont{Manthe}},\ and\ \bibinfo {author}
  {\bibfnamefont{L.}~\bibnamefont{Cederbaum}},\ }%
  \bibfield{journal}{%
  \bibinfo {journal} {Chem. Phys. Lett.}\ }%
  \textbf{\bibinfo {volume} {165}},\ \bibinfo {pages} {73 } (\bibinfo {year}
  {1990}),\ ISSN \bibinfo {issn} {0009-2614}%
  \bibAnnoteFile{NoStop}{MMC90}%
\bibitem{WMM01}%
  \BibitemOpen
  \bibfield{author}{%
  \bibinfo {author} {\bibfnamefont{H.}~\bibnamefont{Wang}}, \bibinfo {author}
  {\bibfnamefont{D.~E.}\ \bibnamefont{Manolopoulos}},\ and\ \bibinfo {author}
  {\bibfnamefont{W.~H.}\ \bibnamefont{Miller}},\ }%
  \bibfield{journal}{%
  \Doi{10.1063/1.1402992}{\bibinfo {journal} {The Journal of Chemical
  Physics}}\ }%
  \textbf{\bibinfo {volume} {115}},\ \bibinfo {pages} {6317} (\bibinfo {year}
  {2001})%
  \bibAnnoteFile{NoStop}{WMM01}%
\end{thebibliography}

\providecommand{\noopsort}[1]{}\providecommand{\singleletter}[1]{#1}%
\end{document}